\documentclass[ ]{article}%
\usepackage[english]{babel}
\usepackage{epsfig}
\usepackage{amssymb}
\usepackage{amsmath}
\usepackage{amsfonts}
\usepackage{graphicx}%
\providecommand{\U}[1]{\protect\rule{.1in}{.1in}}
\begin{document}

\title{ A solution of LIDAR problem in double scattering approximation}
\author{Alexey Buzdin, Sergey Leble*\\
Immanuel Kant Baltic Federal University, A. Nevskogo st. 14
 \\  )* now: Gdansk University of Technology,\\ ul. Narutowicza 11/12, 80-952, Gdansk, Poland,
leble@mif.pg.gda.pl}
\maketitle
\date{\today}

\begin{abstract}
A problem of monoenergetic particles pulse reflection from half-infinite stratified medium is considered in conditions of elastic scattering  with absorbtion account. The theory is based on multiple scattering series solution of Kolmogorov equation for one-particle distribution function.  The analytical representation for first two terms are given in compact form for a  point impulse source and cylindric symmetrical detector.
This publication is an authorized translation (by S. Leble) of a bad-published paper
from 20.06.1980 (available in its original version:
 http://www.mif.pg.gda.pl/krrizm/page/leble/publications.html,
140. A. Buzdin, S. Leble
Lidar Problem Solution in Double-Scattering Approximation
VINITI N2536-80dep (1980) 126 [PDF]). Reading recent articles on the LIDAR sounding of
 environment
 (e.g. Atmospheric and Oceanic
 Optics (2010) 23: 389-395,Kaul, B. V.; Samokhvalov, I. V.) one recovers standing interest to
the related direct and inverse problems. A development of the result for the case of n-fold
scattering and polarization account as well as correspondent convergence series problem
solution of the Kolmogorov equation will be published in nearest future.

\end{abstract}

\section{Introduction}
A problem of monoenergetic particles pulse reflection from half-infinite stratified medium is considered. The particles scattering  is supposed elastic, its interaction is negligible. Such problem relates a LIDAR probing of atmosphere. It is supposed that such process is described by Kolmogorov equation \cite{KolUch} for one-particle distribution function.

The expansion of the distribution function with respect to scattering multiplicity allows to obtain a closed expression for the series terms as multiple integrals.

The main purpose of this paper is a simplification of analytic expressions and investigation of its applicability range in terms of geometric parameters and macroscopic cross-sections. The point impulse source and cylindric symmetrical detector is considered.

Computations by Monte-Carlo method  show a quick convergence of the scattering multiplicity expansion hence the one- and double scattering contributions are enough for the practical LIDAR response evaluation.

\section{The problem formulation}
The equation for the probability density $f=f(t,\vec r,\vec v,t')$ has the form:
\begin{equation}\label{E:con_CNLSE_2_1}
 \frac{1}{c}[\partial_{t'}f+\vec v\cdot\nabla f]=-\sigma_t(z)
f-\int\sigma(cos\gamma, z)f d\Omega,
\end{equation}
Where t'-time, $d\Omega$ - solid angle;  $\sigma - $ bulk differential cross-section of elastic scattering to the angle $\gamma$; $\sigma_t$ - the sum of $\int \sigma d\Omega$ and total cross-section of absorbtion, and
$$\frac{\vec v}{c}=(sin\theta cos\phi,sin\theta sin\phi,cos\theta).$$
In spherical coordinates: $r, \theta$, $\phi$ the scattering angle is expressed as
\begin{equation}\label{cosG}
   cos\gamma =cos\theta cos\theta'+sin\theta sin\theta'cos(\phi-\phi').
\end{equation}
We suppose that the scattering is elastic, $|\overline v|$ does not change while the scattering process occur.
Initial conditions are represented by distributions
\begin{equation}\label{ini}
f(0,x,y,z) = V \delta(x) \delta(y) \delta(z)\delta(\theta).
\end{equation}
It means that we built a solution for the probability density as a weak limit (when $t \rightarrow 0$) to $\delta - function$ at $t>0$.
The distribution $\delta(\theta)$ is chosen as
\begin{equation}\label{deltateta}
    (\delta(\theta),\psi(\theta,\phi))=\int_0^{2\pi}\psi(0,\phi)d\phi.
\end{equation}

\section{Solution}
Let us denote $t=ct'.$
A solution is searched as an N-fold scattering expansion
\begin{equation}\label{U}
    f=f_0+f_1+f_2+...
\end{equation}
We choose for $f_0$,
\begin{equation}\label{Ueq}
 L f_0=\frac{ \partial f_0}{\partial t}+sin\theta cos\phi\frac{\partial f_0}{\partial x}+sin\theta sin\phi\frac{\partial f_0}{\partial y}+cos \theta \frac{\partial f_0}{\partial z}=-\sigma_t(z) f_0,
\end{equation}
and initial condition,
\begin{equation}\label{ini0}
  f_0(0,x,y,z) = V\delta(x)\delta(y)\delta(z)\delta(\theta).
\end{equation}
 For $n\geq1$, the expansion coefficients are defined by
\begin{equation}\label{fn}
    L f_{n+1}=-\sigma_1(z)f_{n+1}+\int_0^{2\pi}\int_0^{\pi}\sigma_2(cos\gamma, z)f_{n}sin\theta'd\theta'd\phi'.
\end{equation}
with initial conditions for $n>0$
\begin{equation}\label{in}
    f_{n+1}|_{t=0}=0.
\end{equation}
Such expansion is usually named as expansion by folds scattering. The expression for $f_0$ is easily found:

\begin{equation}
 \nonumber  f_{0}=\frac{1}{2\pi}\delta(x)\delta(y)\delta(z-t)\delta(\theta)\exp[-\int_0^t\sigma_{t}(z-\tau)d\tau].
\end{equation}
Let us denote a function $E$ via
\begin{equation}\label{E}
    E (t,z,\theta)=\exp[-\int_0^t\sigma_t(z-\tau cos\theta)d\tau],
\end{equation}
then for $f_{n+1}$ we obtain the reccurence
\begin{equation}\label{gen}
\begin{array}{c}
  f_{n+1}(x,y,z,\theta,\phi,t)=\\
\int_0^t E(\tau,z,\theta)\int_0^{2\pi}\int_0^{\pi}\sigma(cos\gamma,z-\tau cos\theta)\\f_n(x-\tau cos\theta cos\phi,y-\tau sin\theta sin\phi,z-\tau cos\theta,\theta',\phi',t-\tau)sin\theta'd\theta'd\phi'd\tau.
\end{array}
\end{equation}
Particulary, for the $n=1$, it yields
\begin{align*}
f_{1}=
\int_0^t E(\tau,z,\theta)E(t-\tau,z-\tau cos\theta,0)\sigma(cos\theta ,z-\tau cos\theta)\\\delta(x-\tau sin\theta cos\phi)\delta(y-\tau sin\theta sin\phi)\delta(z-\tau cos\theta-(t-\tau))d\tau\end{align*}
and for $n=2$
\begin{equation}\label{f2}
\begin{array}{c}
f_{2}=
\int_0^t E(\tau_2,z,\theta)\int_0^{2\pi}\int_0^{\pi}\sigma(cos\gamma,z-\tau_2 cos\theta)\int_0^{t-\tau_2} E(\tau_1,z-\tau_2 cos\theta,\theta')\\E(t-\tau_2-\tau_1,z-\tau_2 cos\theta-\tau_1 cos\theta',0)\sigma(cos\theta' ,z-\tau_2 cos\theta-\tau_1 cos\theta')\\\delta(x-\tau_2 sin\theta cos\phi-\tau_1 sin\theta' cos\phi')\delta(y-\tau_2 sin\theta sin\phi-\tau_1 sin\theta' sin\phi')\\\delta(z-\tau_2 cos\theta-\tau_1 cos\theta'-(t-\tau_2-\tau_1))d\tau_1sin\theta'd\theta'd\phi'd\tau_2.\\
\end{array}
\end{equation}
It is convenient to rerpresent the integrands as distributions in $x,y,z$ space
parametrized by $\theta, \phi,\tau_1, \tau_2, \theta_1, \phi_1 $.
Integrations by $\tau_1, \tau_2, \theta_1, \phi_1$ is understood as integration of the distribution by these parameters. For example, $f_1$ acts as on a function $\psi$ from Shwartz space as
\begin{align*}
(f_1(t,x,y,z,\theta,\phi),\psi(x,y,z))=\\ \int_0^t E(\tau,\tau cos\theta+t-\tau,\theta)E(t-\tau,t-\tau,0)\sigma(cos\theta ,t-\tau)\\\psi(\tau sin\theta cos\phi,\tau sin\theta sin\phi,\tau cos\theta+t-\tau)d\tau.
\end{align*}

\section{Number of particles rate}
Our next problem is evaluation of number of particles which enter the round area of radius $\rho_0$ laying in the plane $xy$ with center in the origin and having velocity vectors inclined to z-axis within the angle interval $\theta\in [\pi-\theta_0,\pi]$.  The angle $\theta_0$ relates the aperture angle of a receiver ( e.g. lidar telescope window). By its direct sense the number is proportional to
\begin{equation}\label{I}
I(t)=\lim_{\Delta t}\frac{1}{\Delta t}\int_{\pi-\theta_0}^{\pi} \int_0^{2\pi}(f(x,y,z,\theta,\phi,t),\psi(x,y,z))sin\theta d\phi d\theta,
\end{equation}
where $\psi(x,y,z)=1 $ for internal points of the domain $x^2+y^2 \leq \rho_0^2,$ \qquad $0\leq z \leq \Delta t |cos\theta|$ and zero outside.

Contribution of $n-fold$ scattering $I_n(t)$ is obtained by substitution $f_n$  instead of $f$ in the Eq. (\ref{I}). Let us evaluate one- and two-fold scattering.
\begin{equation}\label{I1}
\begin{array}{c}
I_1(t)=\\ \lim_{\Delta t}\frac{2\pi}{\Delta t}\int_{\pi-\theta_0}^{\pi} \int_0^{2\pi}E(\tau,\tau cos\theta+t-\tau,\theta)E(t-\tau,t-\tau,0)\sigma(cos\theta ,t-\tau)\\\psi(\tau sin\theta,0,\tau cos\theta+t-\tau)d\tau sin\theta d\phi d\theta.
\end{array}
\end{equation}
The independence of the integrand on $\phi$ is taken into account.

In the problems of LIDAR probing the aperture angle is small $\theta_0<< 1, $ therefore we estimate $cos \theta$ as $-1$. Introduce new variables $z=t-\tau, \xi=\tau sin\theta.$ Intersection of the
 integration domain in (\ref{I1}) with the area in which $\psi$ is nonzero, satisfy the following inequalities (in new variables)
 \begin{equation}\label{ineq}
\begin{array}{c}
\xi \leq \rho_0, \\ 0 \leq 2z-t \leq \Delta t,  \\ 0 \leq \xi \leq \epsilon ( t-z),
\end{array}
\end{equation}
 where $\epsilon=tg\theta_0.$ When $\frac{t}{2}>\frac{\rho_0}{\epsilon}$, the third inequality from (\ref{ineq}) follows from the first ones, hence
 \begin{equation}\label{I11}
\begin{array}{c}
I_1(t)=\\ \lim_{\Delta t\rightarrow 0}\frac{2\pi}{\Delta t}\int_{t/2}^{\frac{t+\Delta t}{2}} \int_0^{\rho_0}E(t-z,2z-t,\pi)E(z,z,0)\sigma(-1,z) \frac{\xi d\xi dz}{(t-z)^2}=
\end{array}
\end{equation}
\begin{equation}\label{F}
\frac{ 2\pi\rho_0^2}{t^2} E(\frac{t}{2},0,\pi)E(\frac{t}{2},\frac{t}{2},0)\sigma(-1,\frac{t}{2}), \end{equation}
which results in LIDAR formula
\begin{equation}\label{F}
\frac{ 2\pi\rho_0^2}{t^2} \exp[-\int_0^{t/2}\sigma_t(z)dz]\sigma(-1,\frac{t}{2}).
\end{equation}
Contribution of the two-fold scattering (\ref{f2}) gives
\begin{equation}\label{I2}
\begin{array}{c}
I_2(t)=\\ \lim_{\Delta t\rightarrow 0}\frac{4\pi}{\Delta t} {\int}^{\pi}_{\pi-\theta_0} \int_0^{\pi}\int_{0}^{t} \int_{0}^{t-\tau}\int_0^{\pi}\textbf{E}\sigma(cos\gamma ,\tau_1(cos\theta_1-1)+t-\tau_2) \sigma( cos\theta_1, +t-\tau_1-\tau_2)\\ \psi(\tau_1sin\theta_1+\tau_2sin\theta_2cos\phi,\tau_2sin\theta_2sin\phi,\tau_1cos\theta_1+\tau_2cos\theta_2+t-\tau_1-\tau_2)   d\theta_1d\tau_2d\tau_1d\phi d\theta_2
\end{array}
\end{equation}
where
\begin{equation}\label{EE}
\begin{array}{c}
\textbf{E}=\exp[-\int_{0}^{t-\tau_2-\tau_1}\sigma_t(t-\tau_2-\tau_1-\tau_1')]d\tau'_1\int_{0}^{\tau_1}\sigma_t((\tau_1-\tau')cos\theta_1+t-\tau_2-\tau_1)d\tau'-\\\int_{0}^{\tau_2}\sigma_t(\tau_1 cos\theta_1+(\tau_2-\tau') cos\theta_2+t-\tau_2-\tau_1)d\tau'
\end{array}
\end{equation}
It was used that the integrand depends only on $\phi=\phi_1-\phi_2$ and, due to definition of $\psi$ and $cos\gamma$, is even in $\phi$.

Let us introduce new variables:
\begin{equation}
\begin{array}{c}
z_1=t-\tau_2-\tau_1,\\
z_2= \tau cos\theta_1 +t-\tau_1-\tau_2,\\
\xi_=\tau_1sin\theta_1,\\
\eta = \tau_2sin\theta_2,\\
\rho=[\tau_1^2sin\theta_1^2+2\tau_1\tau_2sin\theta_1sin\theta_2 +\tau_2^2sin\theta_2^2]^{1/2}
	\end{array}
\end{equation}
The sense of $z_{1,2},\rho$ has transparent intuitive meaning: $z_{1,2}$ are heights of scattering points, $\rho$ is the polar coordinate of a trajectory end. In the domain of consideration there is a one-to-one correspondence between old and new coordinates. The domain is fixed by 	$\xi \geq 0,\eta \geq 0$, and
\begin{equation} \label{14}
\begin{array}{c}
  \tau_1=\sqrt{(z_1-z_2)^2+\xi^2}, \quad \tau_2=t-z_1-\tau_1,\\ cos\theta_1=\frac{z_2-z_1}{\sqrt{(z_1-z_2)^2+\xi^2}},\\ sin\phi=\frac{\sqrt{(\rho^2-(\xi-\eta)^2)(\rho^2-(\xi+\eta)^2)}}{2\xi\eta}
  	\end{array}
\end{equation}

The Jacobian evaluation gives:
\begin{equation}
	Jsin\theta_1sin\theta_2=\frac{\rho}{(\tau_1\tau_2)^2|cos\theta_1|sin\phi},
\end{equation}
that is obtained via transformation formulas. Below, as in derivation of the one-fold scattering, we shall approximate $cos\theta_1$ as 1.

Let us obtain conditions that restrict the integration domain taking into account the definition og $\psi$. The conditions for $\rho$ are obvious. From the problem statement it follows also that $z_1\geq 0, z_2 \geq 0$. Due to $|cos\phi|\leq 1$,
\begin{equation}
	|\xi-\eta|\leq \rho, \quad 	|\xi+\eta|\geq \rho.
\end{equation}
From the choice of $\psi$ and formulas above it follows, that
\begin{equation}\label{17}
	0\leq z_2-t+z_1+\sqrt{\xi^2+(z_1+z_2)^2}\leq\Delta t.
\end{equation}
While $\pi-\theta_c\leq\theta_2\leq\pi,$
\begin{equation}
	0\leq \frac{\eta}{z_2}\leq\epsilon=tg\theta_c.
	\end{equation}
Besides that, the inequalities $0\leq\theta_2\leq t-\tau_1, \quad 0\leq\theta_1\leq t$ holds authomatically because $\tau_1+\tau_2=t-z_1\leq t$ as $\tau_2\geq 0,$ and $z_1\geq =1$. Due to  (\ref{17}),

\begin{equation}\label{19}
	\begin{array}{c}
	0\geq z_1\geq \frac{t+\Delta t}{2} ,\\
	0\geq z_2\geq \frac{t+\Delta t}{2}.
	\end{array}
\end{equation}
Inequality (\ref{17}) with fixed $z_1$ and $z_2$ gives restrictions for the variable $\xi$.
\begin{equation}
	\begin{array}{c}
	\xi_{min}^2=(t-2z_1)(t-2z_1)\geq \xi^2\geq (t-2z_1)(t-2z_1)+\\
	2(t-z_2-z_1) \Delta t+(\Delta t)^2=\xi_{max}^2.
	\end{array}
\end{equation}
Cross-section of area of integration by the plane parallel to the plane $z_1=z_2=\xi=0$ is presented on the Fig. 1 in the case of $\xi< \rho_0,\quad \epsilon z_2 > \xi+\rho_0.$ It is seen that the straight line $\eta=\epsilon z_2$ in dependence on $ z_2, \xi, \epsilon $ crosses shaded area or not. It determines the integral (\ref{I2}) evaluation.

\section{Evaluation and approximate formulas}

 Expression for $I_2(t)$ splits to three contributions $I_{21}, I_{22}, I_{23}$; $I_{22}, I_{23}$ respond the case $\epsilon z_2\leq \xi_{max}+\rho_0 $. For the integral $I_{22}$  $\rho_0\leq\xi_{max},$ while for the $I_{23}$ an opposite equality $\xi_{max}\leq\rho_0 $ holds.

One can show that in conditions of $\epsilon \sigma_{max}\rho_0 ln\frac{t}{\rho_0} << 1 $ and $\frac{t}{2}>>\frac{\rho_0}{\epsilon}$, where $\sigma_{max} = \max_{\gamma} \sigma(cos\gamma)$, integrals $I_{22}$ and $I_{23}$ give contributions small compared to one of one-fold scattering. In particuolar even rough estimation gives:
\begin{equation}
I_{22} \leq \frac{8\pi^2\sigma_{max}^2\rho_0^3\epsilon}{t^2},	
\end{equation}
\begin{equation}
I_{23} \leq	\frac{16\pi^2\sigma_{max}^2\rho_0^3\epsilon}{t^2}(2+\pi+ln\frac{\epsilon t}{2\rho_0}).
\end{equation}
The integral $I_{12}$ in the same conditions is comparable with the contribution of one-fold scattering. Limits of integration in it are found from the condition $\epsilon z_2>\xi_{max}+\rho_0.$ This condition determines the part of the square (\ref{19}), points of which satisfy inequality
\begin{equation}
	z_{1} \geq \frac{t}{2}-\frac{(\epsilon z_2-\rho_0)^2}{2(t-2z_2)}+O(\Delta t).
\end{equation}
 Denoting by the simbol $D_{\Delta t}$ the integration domain of the plane $z_1z_2$, one has
\begin{equation}
\begin{array}{c}
	I_{21}=
	\lim_{\Delta t\rightarrow 0}\frac{4\pi}{\Delta t}
	\int_{D_{\Delta t}}dz_1dz_2\int_{\xi_{min}}^{\xi_{max}}d\xi\int_0^{\rho_0}d\rho\int_{|\xi-\rho|}^{|\xi+\rho|} d\eta
\\	 \frac{\emph{\textbf{E}}\sigma(cos\theta_1,z_1)\sigma(cos\gamma,z_2)2\rho\xi\eta}{((z_1-z_2)^2+\xi^2)(t-z_1-\sqrt{(z_1-z_2)^2+\xi^2})(\rho^2-(\xi-\eta)^2)(\rho^2-(\xi+\eta)^2)}.
	\end{array}
\end{equation}
where $\textbf{E}$ is defined above (\ref{EE}), $cos\theta_1$ is given by (\ref{14}). If one suppose that in a range of small $cos\gamma$ (scattering angles close to $\frac{\pi}{2}$) the cross-section $\sigma$ changes not very quickly, the expression $\sigma(cos\gamma) $ may be approximated by $\sigma(cos(\pi-\theta_1)) $, or, as in \cite{KaSa} , by  $\sigma(cos(\pi-\theta-\theta_0/2)) $.

In such conditions the integration with respect to $\eta,\rho,\xi$ can be performed in explicit form.
After transition to limit of $\Delta t \rightarrow 0, \qquad cos \theta_1=\frac{z_1-z_2}{t-z_1-z_2}$ and the integral under consideration takes the form
\begin{equation}\label{20}
I_{21}=2\pi^2\rho_0^2\int_{D_0}dz_1dz_2\frac{\textbf{E}\sigma(-cos\theta_1,z_2)\sigma(cos\theta_1,z_1)}{z_1^2(t-z_1-z_2)},
\end{equation}
where $D_0$ is defined by following inequalities
\begin{equation}
	0\leq z_1\leq \frac{t}{2},\, 0\leq z_2\leq \frac{t}{2}, z_1\geq \frac{t}{2}-\frac{(\epsilon z_2-\rho_0^2)}{2(t-2z_2)}.
\end{equation}
In variables $z_1,z_2$ the exponential factor is:
 \begin{equation}
	 \textbf{E}=\exp[-2\int_0^{z_1}\sigma_t(\tau)d\tau-\frac{t-2z_1}{t-z_1-z_2}\int_{z_1}^{z_2}\sigma_t(\tau)d\tau ],
\end{equation}
The expression (\ref{20}) is the main result of the article: it represents, in the assumptions made here,
the two-fold contribution into a response of the LIDAR probing.

\section{Conclusion}
The expansion we use converges because the number of particles conserwation law in elastic collisions.
Estimation of  of the convergence speed we delivered for a homogeneous layer  thickness  $d$. In this case the expansion by scattering multiplicity is the power series by optic thickness $\sigma d$.

Consideration of the three-fold contribution into a response, obtained by analogeous scheme in the case of $\sigma_{max} d <1 $ allows to establish conditions in which one can restrict himself by two-terms of the expansion. In this case, in principle, the inverse problem of the function $\sigma$ reconstruction may be posed.

The polarisation account may be held by the scheme we use and do not introduce principal complications.

A diffusion of the mean phase and interferention also may be included in this apparatus and do not contradict simplifications procedure.

The theory development include more detailed modeling of the scattering processes and geometry conditions.

By the way the condition $\sigma_{max} d <1 $ is not necessary in a case of non-spherical indicatrix which is met in real LIDAR problems. The case of strongly nonspherical indicatrix, perhaps, can be considered on a way of unification of our results with ones of \cite{3}.

{\large In this work we present relatively simple method for It means that we
can control evaluation process. }

\end{document}